\documentclass[a4paper,10pt]{article}
\usepackage{amsmath}
\usepackage{amssymb}
\usepackage{graphicx}  
\newcommand{\bea}{\begin{eqnarray}}  
\newcommand{\eea}{\end{eqnarray}}  
\newcommand{\be}{\begin{equation}}  
\newcommand{\ee}{\end{equation}}  
  
\newcommand{\rt}[1]{{}}  
\newcommand{\im}{\mathrm{i}\,}    
\newcommand{\rme}{\mathrm{e}}
\date{}
\begin{document}    
\title {\bf Semiclassical theory of charged pion radiation by nucleons in a strong homogeneous magnetic field }
\author{T. Herpay\footnote{Research Group for Statistical and Biological Physics of the Hungarian Academy of Sciences;  e-mail: herpay@complex.elte.hu} \and A. Patk{\'o}s$^{*}$\footnote{Department of Atomic Physics, E{\"o}tv{\"o}s University, H-1117 Budapest, Hungary; e-mail: patkos@ludens.elte.hu}}


\maketitle


\begin{abstract}
Charged pions produced in very high energy hadronic interactions might be the dominant source of cosmic neutrinos in the GeV--TeV range. Spectral energy power of $\pi^+$ radiation by high energy protons moving in strong magnetic fields typical for magnetars is determined with a semiclassical treatment of the effective pion-nucleon model. The main characteristics emerging from a saddle point approximation to the summation over the allowed range of Landau levels is a sharp lower cut: $E_\pi\ge 0.25 E_p$. The magnitude of the spectral power agrees in this region with the synchrotron radiation spectra of neutral pions.
\end{abstract}

\section{Charged pions as cosmic sources of high energy neutrinos}
Identification of the cosmic sources of (ultra)high energy neutrinos
will make a big leap forward with the completion of the construction
of ICECUBE\cite{icecube06}, 
KM3NeT\cite{km306} and other observatories specifically designed for
neutrino observations, and also with the cosmic ray observatory
P. Auger and its northern counterpart. At present theoretical
arguments are put forward for neutrino production in gamma ray 
bursts\cite{bahcall00}, magnetars 
\cite{zhang03,arons03}, blazars\cite{atoyan01}, pulsars\cite{blasi00} and
compact binary objects\cite{kappes07}. In general
objects where protons might be accelerated to high enough energy
to reach the threshold of pion production in various elementary
reactions are worth to be investigated. High energy neutrinos  are
expected to arise from pion and subsequent muon decays. 

The typical reaction chain in presence of hard enough $\gamma$ quanta
is\cite{waxman97} 
\be
p+\gamma\rightarrow \Delta^+\rightarrow
 n+\pi^+\rightarrow n+\mu^++\nu_\mu\rightarrow n+
\nu_\mu+\nu_e+e^++\bar{\nu}_\mu.
\ee
Energetic
 $\gamma$ quanta might arise from synchrotron radiation of  fast protons
moving on circular orbit in strong magnetic field of magnetars, for
instance. 
In the estimation of the neutrino luminosity possible
 effects on proton acceleration arising from the presence of strong
 magnetic field were taken into account. 
Another mechanism proposed for $\pi^+$ production
involves high energy proton-neutron collisions, assuming that
 also  
the latter component is present in the environment of a gamma ray burst\cite{bahcall00}
\be
p+n\rightarrow n+n+\pi^+\rightarrow 2n+\mu^++\nu_\mu\rightarrow
2n+e^++\nu_e+\bar\nu_\mu+\nu_\mu.
\ee
The problem of neutral pion production as a result of synchrotron
radiation of pion field strongly coupled to the protons
in presence of strong magnetic field $B$: 
\be
p+B\rightarrow p+\pi^0+B
\ee
was investigated first 
forty years ago by Ginzburg and Zharkov \cite{ginzburg64} and reviewed by 
Ginzburg and Sirovatski\cite{ginzburg65}. The treatment followed
closely the classical theory of synchrotron radiation. It was shown to
reproduce the results of first order quantum perturbation theory
applied to an effective nucleon-pion interaction Hamiltonian. 
A perturbative estimate for
the $p\rightarrow n+\pi^+$ transition probability under 
the influence of a strong electromagnetic wave 
(i.e. charged particles considered as Volkov states) 
was given at the same time by
Zharkov\cite{zharkov65}.  

At present much higher cosmic magnetic fields
are known, in particular in magnetars, where $B=10^{14-16}$G was
observed\cite{harding06}. Substantial progress has been achieved also 
in the understanding
of acceleration mechanisms. Therefore it is timely to evaluate the old
formulae with the actual parameters. This task was partially done by
Tokuhisa and Kajino\cite{tokuhisa99}, but no reassessment was 
attempted to date for the case of charged pions.
  
The subject of this note is to present a semiclassical theory for
charged pion radiation in strong magnetic field:
\be
p+B\rightarrow n+\pi^++B
\label{ch-pi-B}
\ee
 and estimate the luminosity of this radiation
for the highest possible energy range. Since any individual proton
contributes only a single positively charged pion into the radiation, 
it is hard to
associate a classical charged pion source with a single proton. If one
assumes a very intense proton beam, however, then the loss of a
restricted number of
protons should not mean a strong back reaction, e.g. the source
remains approximately intact. This kind of approximation represents
 the main hypothesis of this work, which assumes unchanged
proton source 
during the building up of a classical $\pi^+$ radiation field around
the 
source. 
\section{The model}
The Lagrangian of charged pions coupled to a nuclear source density is
the following:
\be
  \mathcal{L} =(\partial_\mu -\im e A_\mu)\pi^+(\partial_\mu +\im e
A_\mu)
\pi^--m^2\pi^+\pi^--g\sqrt{2}(\rho_N^-\pi^++\rho_N^+\pi^-)\,,
\ee
where the nuclear source terms are $\rho_N^-=\im\bar{p}\gamma_5n$,
$\rho_N^+=\im\bar{n}\gamma_5 p$, and $g^2\approx 14.6$. (Throughout this paper the unit system
$\hbar=c=1$ is used, SI values for some important dimensional
combinations will be explicitly given below). 
The Euler--Lagrange equations of the pion fields, in
Coulomb gauge ($A_0=0,\,\partial_i A_i=0$) 
read as
\be
\left[\partial_t^2-\partial_i\partial_i\pm 2\,\im e A_i\partial_i+e^2
  A_i 
A_i+m^2\right]\pi^\pm=- g \sqrt{2}\rho^\pm_N\,.\label{e:EL1}
\ee
In presence of a static magnetic field along the '$z$' direction 
($\mathbf{B}=(0,0,B)$), it is useful to introduce cylindrical coordinates.
Then (\ref{e:EL1}) takes the following form:
\be
  \left[\partial_t^2-\frac{1}{\rho}\frac{\mathrm{d}}{\mathrm{d}\rho}
\left(\rho\frac{\mathrm{d}}{\mathrm{d}\rho}\right)+\frac{1}{\rho^2}
\frac{\mathrm{d^2}}{\mathrm{d}\phi^2}+\frac{\mathrm{d^2}}{\mathrm{d}z^2}\pm
  \im e B
  \frac{\partial}{\partial\phi}+\frac{1}{4}e^2B^2\rho^2+m^2\right]\pi^\pm=- 
g \sqrt{2}\rho^\pm_N\,.
\label{e:EL2}
\ee
The particular retarded solution describing the radiation of charged
pions for the above linear partial
differential 
equations is constructed as 
\be
  \pi^\pm(\rho,\phi,z,t)=-g \sqrt{2} \int\mathrm{d}\rho^\prime\mathrm{d}
\phi^\prime\mathrm{d}z^\prime\mathrm{d}t^\prime G^\pm(\rho,\phi,z,t;
\rho^\prime,\phi^\prime,z^\prime,t^\prime) \rho_N^
\pm(\rho^\prime\phi^\prime,z^\prime,t^\prime),\label{e:sol}
\ee
where the Green's functions $G^\pm$ are determined by the eigenvalues 
and eigenfunctions of the differential operator of the left hand side 
of (\ref{e:EL2}),
\be
  G^+ (\rho,\phi,z,t;\rho^\prime,\phi^\prime,z^\prime,t^\prime)=\sum_i 
\frac{\Phi_i(\rho,\phi,z,t)\Phi_i^\star(\rho^\prime,\phi^\prime,
z^\prime,t^\prime)}{\lambda_i}=(G^-)^\star .\label{e:greendef}
\ee
Writing the eigenfunctions in the form
\be
\Phi=\mathrm{exp}\left[\im(\mu\phi+kz-\omega t)\right]R(x)\,,\qquad 
x=\frac{eB}{2}\rho^2\,,\label{e:ef}
\ee
the corresponding eigenvalue problem is reduced to the eigenvalue 
problem of an ordinary differential operator,
\be
  xR^{\prime\prime}+R^\prime-\left[\frac{x}{4}+\frac{\mu^2}{4x}-
\frac{\mu}{2}-\frac{1}{2eB}\left(\omega^2-k^2-m^2\right)\right]R(x)
=-\lambda R(x).
\ee
This is the well-known problem of the relativistic Landau-levels, 
with the normalized  solution (see for example in \cite{abr} 
22.6.16 and 22.2.12)
\be
  R_n^\mu(x)\!=\!\sqrt{\frac{\Gamma(n+1)}{\Gamma(n+|\mu|+1)}}\,\rme^{-x/2} x^{|\mu|/2}L_n^{|\mu|}(x),\,
\lambda=eB(2n+|\mu|-\mu+1)-\omega^2+k^2+m^2,\label{e:rnmu}
\ee
where $L_n^{|\mu|}(x)$ denotes the generalized Laguerre polynomial,
and  $R_n^\mu(x)$ form a complete orthogonal set for $n=0,1,2,3,\ldots$ .
The normalized eigenfunctions (\ref{e:ef}) are written in the
following explicit form:
\be
\Phi_{n,\mu,k,\omega}(\rho,\phi,z,t)=\sqrt{\frac{eB}{(2\pi)^3}}\,
{\exp}\left[\im(\mu\phi+kz-\omega t)\right]
 R_n^\mu\left(\frac{eB}{2}\rho^2\right)
\ee
and the corresponding  eigenvalues are listed in (\ref{e:rnmu}). 
From (\ref{e:greendef}) the Green's function is found after 
performing the $\omega$ integration:
\bea
  G^+ (\rho,\phi,z,t;\rho^\prime,\phi^\prime,z^\prime,t^\prime)&=&
\frac{eB}{(2\pi)^2}\int\mathrm{d}k\sum_{n=0}^{\infty}\sum_{\mu=
-\infty}^{\infty} {\exp}
\left[\im(\mu(\phi-\phi^\prime)+k(z-z^\prime))\right]\nonumber\\
  &\times& R_n^\mu(\rho) R_n^\mu(\rho^\prime)
\frac{1}{A_{n,\mu,k}}\sin\left[A_{n,\mu,k}\left(t-t^\prime\right)\right]
\theta(t-t^\prime),
\eea
where $A_{n,\mu,k}^2=k^2+m^2+eB(2n+|\mu|-\mu+1)>0$.

The source density is proportional to the transition matrix element
from a Landau state of a proton to a plane wave neutron. 
We assume that a macroscopic
circular proton current is present and neglect the decrease of its
intensity due to the
$\pi^+$ radiation. 
The corresponding classical proton density is given by
\be
\rho_N^+(\rho,\phi,z,t)=\sqrt{1-v_p^2}\,\frac{1}{\rho}\delta(\rho-R)\delta(z)
\sum_\nu \frac{1}{2\pi} \rme^{\im\nu(\phi-\omega_st)},
\ee 
where $v_p$ is the velocity of the proton
(${1-v_p^2}=M_p^2/E_p^2=1/\gamma^2$) on the $z=0$ plane (we do not
consider the motion of the proton along the  $z$ axis) and $R$,
$\omega_s$ are the relativistic cyclotron radius and frequency of the
proton. Substituting this source into (\ref{e:sol}) the charged meson 
radiation field is determined after performing the $k$-integration:
\bea 
   \pi^+(\rho,\phi,z,t)&=&\frac{g\sqrt{2}}{\gamma}
\frac{eB}{2\pi}\sum_{n,\mu}\frac{-\im\mathrm{sign\,}(\mu)}{2
\Omega_{n,\mu}} \exp\left[\im(\mu(\phi-\omega_s t)+\mathrm{sign\,}
(\mu z)\Omega_{n,\mu}z)\right]\nonumber\\
  &\times&R_n^\mu\left(\frac{eBR^2}{2}\right)
R_n^\mu\left(\frac{eB\rho^2}{2}\right).
\label{e:rfunct}
\eea
Here $\Omega_{n,\mu}=\sqrt{\mu^2\omega_s^2-m^2-eB(2n+|\mu|-\mu+1)}$. 
The allowed range of the summation is restricted by the positivity 
requirement of the argument of the root. The case of positive and 
negative values of $\mu$ represent two distinct series 
in the sum $\sum_{n,\mu}=\sum^++\sum^-$, where the respective range of 
summation is given as follows:
\begin{equation}
   \sum^+ \left\{\begin{array}{l}
      \sqrt{eB+m^2}< \mu\omega_s\\ 0\le n\le
   \frac{(\mu\omega_s)^2-m^2-eB}{2eB}= n_+\end{array}\!\!\!\,\!\!\right.,\,\,
\sum^-\left\{
  \begin{array}{l}\sqrt{(eB/\omega_s)^2+eB+m^2}+eB/\omega_s<|\mu|\omega_s \\ 
0\le n\le \frac{(\mu\omega_s)^2-2eB|\mu|-m^2-eB}{2eB}=
n_-\end{array}\right.\!\!. \label{e:cond}
\end{equation}
The pion field is localized on Landau levels radially, therefore 
the energy flux is nonzero  only along the $z$ axis, e.g. one has to 
consider only the
'$0z$' component of the energy momentum tensor,
\begin{equation}
T_{0z}=\left[\frac{\partial\pi^+}{\partial
    t}\left(\frac{\partial\pi^+}{\partial z}\right)^*+
\left(\frac{\partial\pi^+}{\partial
    t}\right)^*\frac{\partial\pi^+}{\partial z}\right],
\end{equation}
and the power of the radiation is given by computing the surface 
integrals of this flux on some distant $\pm z_0$ planes:
\be
  P=\int \mathrm{d}\rho\rho\int \mathrm{d}\phi (T_{0z}(z=+z_0)-T_{0z}(z=-z_0))
=\frac{g^2}{\gamma^2}\frac{eB}{\pi}\sum_{n,\mu}
\frac{|\mu|\omega_s}{\Omega_{n,\mu}}
\left[R_n^\mu\left(\frac{eBR^2}{2}\right)\right]^2. \label{e:finalP1}
\ee
\section{The spectral power of the pion radiation}

The classical-quantum correspondence tells that the energy of the
emitted pions is given by $|\mu|\omega_s$. Therefore the $\mu$-sum
should be converted into $E_\pi$-integral if we wish to derive a
formula for the spectral power. For this conversion one has to set a
stage for the actual order of magnitude of the 
characteristic quantities.

The frequency and the radius of the cyclotron trajectory of protons
with energy $E_p$ are
\be
\omega_s=\frac{eB}{E_p}=\frac{1}{\gamma^2}\frac{B}{B_0}E_p,
\quad  R=\frac{v_p}{\omega_s}=\frac{\sqrt{1-
\frac{M_p^2}{E_p^2}}}{eB}E_p=\sqrt{\frac{{E_p^2-M_p^2}}{e^2B^2}}, 
\label{e:omegas}
\ee
where $\gamma\equiv E_p/M_p$ and
$B_0\equiv M_p^2/e$ (in SI units $M_p^2 c^3/(e\hslash)
\approx 1.5\cdot 10^{20}\,$G). In  
(\ref{e:finalP1}) $R_n^\mu$ depends on the combination 
\be
\xi=\frac{eBR^2}{2}=\frac{\gamma^2-1}{2}\frac{B_0}{B}.
\ee
The actual value of $\xi$ for a typical cosmic setting can be
estimated by 
taking the values (in SI units) $B=10^{15}$\,G and $\gamma=10^5$, which yield 
$\xi=7.4\cdot 10^{14}$.

Going from the $\mu$-sum to the $E_\pi$-integration the pion radiation
power  
can be written in the following form (again in $\hbar=c=1$ units):
\begin{eqnarray}
P&=&\frac{g^2eB}{\pi\gamma^2\omega_s}\left(\int_{E_+}^\infty
  \mathrm{d}\,E_\pi \sum^+_{n} \frac{E_\pi\left(R_{n}^{E_\pi/\omega_s} 
(\xi)\right)^2}{\sqrt{E_\pi^2-eB(2n+1)-m^2}} \right.\nonumber\\
&+&\left.\int_{E_-}^\infty
\mathrm{d}\,E_\pi \sum^-_{n} \frac{E_\pi
  \left(R_{n}^{E_\pi/\omega_s}(\xi)
\right)^2}{\sqrt{E_\pi^2-eB(2n+2E_\pi/\omega_s+1)-m^2}} \right),
\end{eqnarray}
where from (\ref{e:cond}) and (\ref{e:omegas})
\begin{equation}
E_+=\sqrt{eB+m^2},\quad
E_-=\sqrt{\left(\frac{eB}{\omega_s}\right)^2+eB+m^2}
+\frac{eB}{\omega_s}>2E_p.
\end{equation}
The  second integral is irrelevant due to
$E_->2E_p$ (in the classical approximation without back reaction the
limitation on the pion energy $E_\pi\leq E_p$ is imposed by hand!). The 
differential power of the radiation per unit pion energy is
given by
\be
\frac{\textnormal{d} P}{\textnormal{d}E_\pi}=\frac{g^2eB}{\pi\gamma^2
\omega_s} \sum_{n=0}^{n^+} \frac{E_\pi
\left(R_{n}^{E_\pi/\omega_s}(\xi)
\right)^2}{\sqrt{E_\pi^2-eB(2n+1)-m^2}}. 
\ee 
Since $\xi$ is a dimensionless (large) parameter, it is useful to express all
quantities in proportion to $\xi$,
\be
  \mu\omega_s=E_\pi\,,\quad \beta=\frac{E_\pi}{E_p}\,,\quad n_+=
\frac{E_\pi^2-m^2}{2eB}-\frac{1}{2},\quad eB=M_p^2\frac{\gamma^2-1}{2\xi},
\ee
\be
  \Delta=\frac{n}{\xi},\,\,\,\, \frac{E_\pi}{\omega_s}=
\frac{E_\pi E_p}{eB}=\frac{E_\pi
  E_p}{M_p^2}\frac{2\xi}
{\gamma^2-1}=2\beta\frac{\gamma^2}{\gamma^2-1}\xi \equiv 2\beta a \xi,\,\,\,
\,(a\approx1).
\ee
With these notations, from  (\ref{e:rnmu}) and  (\ref{e:rfunct}),
\be
\left(R_{n}^{E_\pi/\omega_s}(\xi)\right)^2={\frac{\Gamma(\Delta\xi+1)}
{\Gamma(2\beta a\xi+\Delta\xi+1)}}\exp(-\xi)\xi^{2\beta a\xi}
\left(L^{2\beta a\xi}_{\Delta\xi}\left( \xi\right)\right)^2.
\ee
Since the change in $\Delta$ for one step in $n$
is very 
small,  $\sum_n$ can be written as $ {\xi}\int \textnormal{d} \Delta$,
and the 
following final expression is obtained:
\be
\frac{\textnormal{d}
  P}{\textnormal{d}E_\pi}=\frac{{g^2}}{\pi}\frac{M_p\xi}{\gamma}
\int_{0}^
{\Delta^+} \textnormal{d} \Delta \frac{\beta}{\sqrt{\beta^2-a\Delta-
\frac{m^2/M_p^2+B/B_0}{\gamma^2}}}F^2(\beta,\Delta,\xi), \label{e:integr} 
\ee
where
\be
  F^2(\beta,\Delta,\xi)=\left(R_{n}^{E_\pi/\omega_s}(\xi)
\right)^2,\,\,\,\,
\Delta^+=\frac{\gamma^2-1}{\gamma^2}\left(\beta^2-\frac{m^2/M_p^2+B/B_0}
{\gamma^2} \right).
\ee
The final task is to evaluate this integral. The difficulty is that the 
indices of the generalized Laguerre polynomials both are proportional to 
its argument $\xi$, which takes asymptotically large value. It appears  
somewhat discouraging that the integrand contains the very small  
factor $\rme^{-\xi}$. Therefore the  
integration should be performed with great precaution. A saddle point   
representation will be described in the next subsection for the asymptotic  
regime of the Laguerre polynomials and shown to cancel  
exactly the suspicious small factor.

\section{Saddle point evaluation of the spectral $\pi^+$ radiation   
power}        
    
The key step in the evaluation of the spectral power is to find the  
appropriate asymptotic representation of $L_{\Delta\xi}^{2\beta  
  a\xi}(\xi)$ and with this for the factor $F(\beta,\Delta,\xi)$ in  
the integrand of (\ref{e:integr}).     
    
\begin{figure}[t]
\begin{center}
\includegraphics[width=0.99\textwidth]{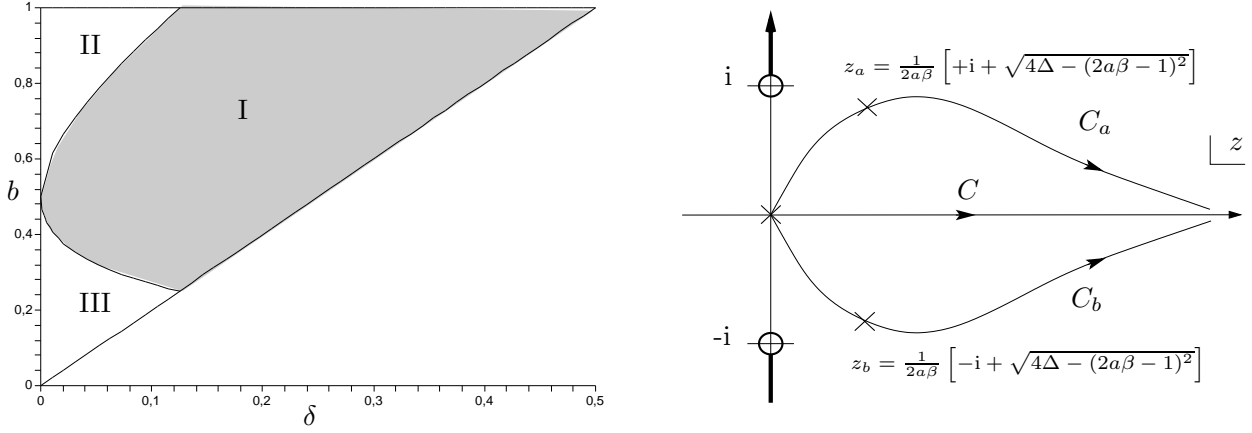}
\caption{On the left: For values of $b=2\beta a$ and
  $\delta=2\beta a \Delta$ in the shaded
  region (I), the saddle points have non-zero real parts. Otherwise
  (in regions II and III) the saddle points are on the imaginary
  axis. The boundary of the shaded area on the right hand side
  ($b=2\delta$) corresponds   to (\ref{e:bound}).
 On the right: The complex saddle points ($0,\,z_a,\,z_b$) of $q_a$
 and $q_b$ when ($b$,\,$\delta$)$\in I$, the original contour along
the positive real axis
  ($C$) and the deformed integration contours ($C_a$,
  $C_b$).\label{f:cont}}
\end{center}
\end{figure}

The generalized Laguerre polynomials can be expressed with help of  an  
 integral of the Bessel functions (see  22.10.14 in \cite{abr}),  
\be  
  L^\mu_n(\xi)=\frac{1}{\Gamma(n+1)}\xi^{\mu/2} \rme^{\xi}\int_0^\infty  
\mathrm{d} k \rme^{-k} k^{n+\mu/2}  
J_\mu(2\sqrt{k\xi})\,,\quad\xi=\frac{\mu} 
{2a\beta}\,.   \label{e:bessel}
\ee  
Using a convenient asymptotic form of the Bessel functions,
$F(\beta,\Delta,\xi)$ ($R^\mu_n(\xi)$) can be written in the following form   
\be
  F(\beta,\Delta,\xi)= N^\prime\int_0^\infty \mathrm{d}z \sqrt{z}\left(\rme^{\mu
      q_a(z)}+\rme^{\mu q_b(z)}\right), \label{e:Fintegrand}  
\ee  
where the expressions of $q_{a,b}(z)$ and $N'$ can be found in the Appendix.
One finds that both terms of the integrand can be  
analytically continued into the complex $z$-plane. One can find its  
complex saddle points and can deform the integration path from the  
real axis into a curve of steepest descent in the complex z-plane which connects the real  
lower and the upper limiting points with the saddles (see the right hand side of figure \ref{f:cont}).  The integral for  
large $\mu(=2\beta a\xi)$ is dominated by the result of the 
Gaussian integration  
around the tallest saddle.  The location of the saddle points can be
given in different ways for  different regions of the ($\beta,
\Delta$) parameter space
(see the left hand side of figure \ref{f:cont}). 
From the detailed saddle
point analysis it can be concluded
that the exponentially small $N^\prime$ factor in (\ref{e:Fintegrand}) 
is cancelled by the contribution
of the saddle point only when  ($\beta$, $\Delta$) are in the region
I.  The interested reader can reconstruct the  
steps of this argument by the details given in the Appendix.      
The result of the saddle point evaluation is the following:
\be  
  F(\beta,\Delta,\xi)=\sqrt{\frac{2}{\pi\xi}}\left[\frac{1}  
{4\Delta-(2a\beta-1)^2}\right]^{1/4}\sin\left(\xi\Phi_1(\beta,\Delta)+  
\Phi_2(\beta,\Delta)+\pi/4\right),\label{e:vegrn}  
\ee  
where  
\bea  
  \Phi_1&=&2\Delta\textrm{arctan}\,\frac{\sqrt{4\Delta-(2\beta a-1)^2}}  {(2a\beta -1)}-(4a\beta+2\Delta)\textrm{arctan}\,\frac{\sqrt{4\Delta  -(2a\beta-1)^2}}{(2a\beta+1)}\nonumber\\  
 &+&\sqrt{4\Delta-(2a\beta-1)^2},\\  
 \Phi_2&=&\textrm{arctan}\,\frac{\sqrt{4\Delta-(2a\beta-1)^2}}{(2a\beta-1)}  
-2\textrm{arctan}\,\frac{\sqrt{4\Delta-(2a\beta-1)^2}}{(2a\beta+1)},  
\eea  
when ($\beta$,\,$\Delta$)$\in$\,I, and $F(\beta,\Delta,\xi)\approx
0$ when ($\beta$,\,$\Delta$)$\in$\,II or III for large $\xi$.

Returning to  the expression of the differential power of the  
radiation per unit pion energy, the factor $F^2(\beta,\Delta,\xi)$
appearing in  
(\ref{e:integr}) is just the square of the expression which appears  
in (\ref{e:vegrn}), therefore  
\bea  
 \frac{\textnormal{d}  
  P}{\textnormal{d}E_\pi}&=&\frac{{g^2}}{\pi}\frac{M_p\xi}{\gamma}  
\int \textnormal{d} \Delta \frac{\beta}{\sqrt{\beta^2-a\Delta-  
\frac{m^2/M_p^2+B/B_0}{\gamma^2}}} \nonumber\\  
 &\times& \frac{1}{\pi\xi}\sqrt{\frac{1}{4\Delta-(2\beta a-1)^2}}\, 
\left[1+\sin\left( 2\xi\Phi_1(\beta a,\Delta)+2\Phi_2(\beta  
    a,\Delta)\right)  
\right]\,  
\eea  
The strongly oscillating term of the integrand would give ${\cal  
  O}(1/\xi)$ contribution therefore one can neglect it and one finds
a simple result which is different from zero for a restricted range of
$E_\pi/E_p$ ($0.25\le E_\pi/E_p\le \gamma^2$, for $\gamma^2\gg1$) :    
\bea  
 \frac{\textnormal{d}  
  P}{\textnormal{d}E_\pi}&=&{\frac{g^2}{\pi^2}}\frac{M_p}{\gamma}   
\int_{(2\beta a-1)^2/4}^{\Delta^+} \textnormal{d} \Delta \frac{\beta}  
{\sqrt{\beta^2-a\Delta-\frac{m_\pi^2/m_p^2+B/B_0}{\gamma^2}}}\frac{1}  
{\sqrt{4\Delta-(2\beta a -1)^2 }} \nonumber\\  
 &\approx&\frac{g^2}{2\pi}\frac{E_\pi}{\gamma^2}, \,\,\,\,\,\,  
\textrm{if}\,\qquad\frac{1}{4}  
\le\frac{E_\pi}{E_p}\le\gamma^2\,.        \label{e:final1}
\eea    
This simple result can be used actually only for $E_\pi<E_p$ since the  
energy of the radiated pion cannot exceed the energy of its parent proton.    
The result is compared next with the results of \cite{tokuhisa99} and  
\cite{ginzburg64}.

\section{Numerical results}

\begin{figure}
\begin{center}
\includegraphics[width=0.78\textwidth]{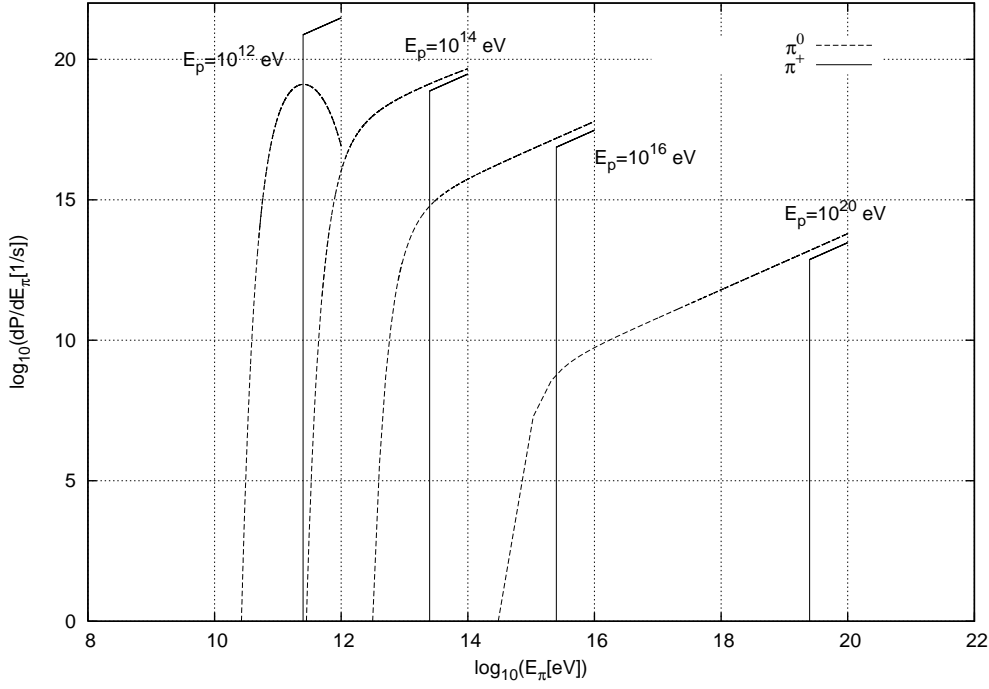}
\end{center}
\caption{The differential radiation power per unit $\pi^+$ energy
  for different proton energies in comparison with the $\pi_0$
  radiation given by \cite{tokuhisa99} for strong  ($B=10^{16}$ G) 
magnetic fields.\label{f:final1}}
\end{figure}

In figure \ref{f:final1} we plotted the above expression and the
corresponding $\pi_0$ spectra for
different proton energies at $B=10^{16}$ G. In this plot we restrict 
the emitted  pion energy to
the physically interesting region ($E_\pi<E_p$) as in \cite{tokuhisa99}. It is
interesting that the width of spectrum in our case is
independent of the proton energy (in logarithmic scale) in contrast to
the case of $\pi_0$ emission. This behaviour is obvious from the very
simple expression (\ref{e:final1}), and this  is the
consequence of the large $\xi$ asymptotic form of the 
Laguerre polynomial (\ref{e:vegrn}),
which becomes exponentially small  
when  $\beta=E_\pi/E_p$ goes over from region I to region
II or III (see figure \ref{f:cont}). The location of the 
intersection  of the
boundary of region I and of the integration range of (\ref{e:final1}) is
independent of $\xi$ (and $B$) and lies at $E_\pi/E_p=1/4$. 
In our approximation the slope of the spectrum variation  is
infinite at $E_\pi/E_p=1/4$  due to the very different behaviour of
the Laguerre polynomials in regions I and III.  As a matter of
fact, our saddle point analysis is not valid in a very small vicinity of the
 boundary of regions I and II.  due to the fact that the
 second derivative of $q_{a,b}$ in (\ref{e:Fintegrand})
tends to zero when $\beta,\,\Delta$ reach the boundary. The detailed
investigation of the saddle point structure of Laguerre polynomials
shows that the width of this ``transition'' region is $\approx 1/\xi$,
and the slope of the spectrum at $E_\pi/E_p=1/4$ in
figure \ref{f:final1} is proportional to $\xi$ (logarithmic scale!).

Another interesting property of the $\pi^+$ radiation is that  the
total emitted pion energy (integral of the expression (\ref{e:final1}) over
$E_\pi$) is independent of the magnetic field  in our approximation. This
$B$ independence is also shown in \cite{tokuhisa99} for the $\pi^0$ 
radiation at large magnetic
fields and/or  proton energy, in contrast with \cite{ginzburg64}, 
where the total
emitted $\pi^0$ energy is proportional to $B^2$. The origin of the difference
between these two results (although both papers are based
on the same semiclassical treatment of the $\pi_0$ radiation, 
calculated in \cite{ginzburg64}) is that in \cite{tokuhisa99} the 
differential power of
radiation is integrated over $E_\pi$ from $m_\pi$ to $E_p$, 
while in \cite{ginzburg64} the
integration range is infinite, and the dominant contribution  of
emitted energy comes from the unphysical ($E_\pi>E_p$) region of the
spectrum in case of large $B$ and/or $E_p$.

\section{Discussion}
The prospective astrophysical interest of the very high energy
neutrino source discussed in this paper depends on the simultanous 
occurrence of very strong magnetic fields and very energetic
(e.g. $100$ TeV) protons. Different propositions put proton
acceleration at different
 distances from the surface of the neutron core, therefore 
the actual magnitude of the magnetic field strength might vary
 strongly. Most optimistic is the so-called ``light-cylinder''
 accelerator mechanism, which in fast rotating objects can be located
 close to the surface. In this case the magnetic field strength is not
 much weaker than  that 
directly measurable on the surface. Since surface keV-photons
 are necessarily present a concurrence between resonant photoproduction
of charged pions (from the decay of $\Delta^+$) and non-resonant
synchrotron radiation-like-production (\ref{ch-pi-B}) should be
envisaged. There could be 10-20 magnetars in our galaxy, with the
great observational advantage that their
activity is not restricted to only a few minutes, like in case of
GRB's. Assuming some well-defined galactic  distribution for these objects,
one combines it with the neutrino luminosity of single objects to
estimate the muon flux generated by $\nu$-detection in ice. 

\section*{Acknowledgements}
The authors are indebted to prof. Peter M{\'e}sz{\'a}ros for suggesting the
subject of this investigation and for patiently explaining its
astrophysical interest. This research was supported by the Hungarian
Research Fund (OTKA-T-046129).

\section*{Appendix}  
We present below a detailed derivation of the asymptotic behaviour of 
 (\ref{e:Fintegrand}) when $\xi>>1$. The notation looks somewhat simpler
if one goes back to the original Landau level wave functions
\be
F(\beta,\Delta,\xi)=R^\mu_n\left(\frac{\mu}{2b}\right), \qquad
b\equiv a\beta.
\ee
Relations between the variables $\Delta=n/\xi,\mu=2b\xi$ were used in
the main text. The integral representation of this function has now
the following form:  
\bea  
R^\mu_n\left(\frac{\mu}{2b}\right)&=&\frac{1}{\sqrt{\Gamma(n+1)  
\Gamma(n+\mu+1)}}\exp\left(\frac{\mu}{4b}\right)  
\left(\frac{\mu b}{2}\right)^{n+\mu/2+1} \nonumber  
\\  
&\times&{\int_0^\infty  
\mathrm{d}x \exp\left(\mu\left[-xb/2+(\delta+1/2)\ln  
    x\right]\right)J_\mu(\mu\sqrt{x})}\,, \label{e:alaprn} 
\eea  
where a new integration variable is introduced instead of the variable 
appearing in  (\ref{e:bessel}) and also a new parameter is defined:  
\be  
k=x\frac{\mu b}{2},  
\qquad 0<\delta\equiv \frac{n}{\mu}<\frac{b}{2}\,. \label{e:bound}  
\ee  
In our application $\mu$ is large,    
one can use the asymptotic form of the Bessel functions given by
 9.3.2 and 9.3.3 of \cite{abr},   
\be  
  J_\mu(\mu\sqrt{x})\!=\!\!\frac{1}{\sqrt{2\pi\mu}}\times\left\{\begin{array}{l}  
\frac{2}{\sqrt{{\tan} K_1}}\cos(\mu(\tan K_1-K_1)-\pi/4),\,\,\,\,\,\textrm{if } x>1\\   
\frac{1}{\sqrt{\tanh K_2}} \exp(\mu (\tanh K_2-K_2)),\,\, \quad\quad  \textrm{if }   
x<1 \end{array}\right.\!\!,  
\ee  
where   
\be  
\begin{array}{l} \cos K_1=\sqrt{1/{x}}, \\\tan K_1=\sqrt{x-1},  
\end{array}  
\qquad  \begin{array}{l} \cosh K_2 =\sqrt{1/{x}}, \\\tanh  
  K_2=\sqrt{1-x}\,,   
\end{array}   
\ee  
Therefore  (\ref{e:alaprn}) is broken up into the sum of two integrals:
\be   
R^\mu_n\left(\frac{\mu}{2b}\right)=N(I_1+I_2),
\ee
where  
\bea  
  I_1&=&\!\int_1^\infty \!\!\mathrm{d} x \frac{2}{(x-1)^{1/4}}  
\exp\left(\mu\left[-\frac{xb}{2}+(\delta+\frac{1}{2})\ln  
    x\right]\right)\nonumber\\  
 &\times&\cos\left(\mu\left[  
    \sqrt{x-1}-\mathrm{acos}\,\left(\frac{1}{\sqrt{x}}\right)\right]-  
\frac{\pi}{4}\right),\label{e:i1} \\  
  I_2&=&\!\int_0^1 \mathrm{d} x \frac{1}{(1-x)^{1/4}}  
\exp\left(\mu\left[-\frac{xb}{2}+(\delta+\frac{1}{2})\ln  
    x+\sqrt{1-x}-\mathrm{acosh}\,\left(\frac{1}{\sqrt{x}}\right)\right]  
\right)\!,\\  
  N&=&\frac{1}{\sqrt{2\pi\mu\Gamma(n+1)\Gamma(n+\mu+1)}}\exp  
\left(\frac{\mu}{4b}\right)\left(\frac{\mu b}{2}\right)^{n+\mu/2+1}.  
\eea  
Moreover, $N$ can be replaced in the asymptotic regime
 using Stirling's formula for the $\Gamma$-functions,  
\bea  
  N&\approx&\frac{1}{2\pi\sqrt{\mu}}\exp\left[\mu\left(\frac{1}{4b}+  
\frac{1}{2}\right)+n+1+\frac{\mu}{2}\ln\left(\frac{\mu b}{2(n+\mu+1)}  
\right)\right.\nonumber\\  
 &+&\left.n\ln\left(\frac{\mu b}{2\sqrt{n+1}\sqrt{n+\mu+1}}\right)+  
\ln\left(\frac{\mu b}{2\left(n+\mu+1\right)^{1/4}  
\left(n+1\right)^{1/4}}\right)\right].\label{e:N2}  
\eea    
Now we can proceed with the evaluation of $I_1$. The method will be
similar also for $I_2$, and will be just stated after the discussion
of the case of $I_1$.    
One changes once more the integration variable in  (\ref{e:i1}),  
\be  
  I_1=2 \exp\left(-\frac{\mu b}{2} \right)\int_0^\infty \mathrm{d} z  
\sqrt{z}   
\left[\exp\left(\mu q_a(z)-\mathrm{i}\frac{\pi}{4}\right)+\exp\left(\mu  
  q_b(z)+\mathrm{i}\frac{\pi}{4}\right) \right],\label{e:i12}  
\ee  
where  one introduces
\be  
x=z^2+1\,\leftrightarrow (x-1)^{1/4}=\sqrt{z}\,,  
\ee  
and  
\bea  
q_a(z)&=&-z^2\frac{b}{2}+\left(\delta+\frac{1}{2}\right)  
\ln(z^2+1)+\mathrm{i}\sqrt{z^2}-\mathrm{i}\,\mathrm{acos}\sqrt{\frac{1}  
{z^2+1}},\\  
q_b(z)&=&-z^2\frac{b}{2}+\left(\delta+\frac{1}{2}\right)  
\ln(z^2+1)-  
\mathrm{i}\sqrt{z^2}+\mathrm{i}\,\mathrm{acos}\sqrt{\frac{1}{z^2+1}}.  
\eea    
In  (\ref{e:i12}) the integrand can be continued analytically
 into the complex  
$z$ plane, which allows the use of 
the method of steepest descent and stationary  
phase for the evaluation of the integral
(saddle point approximation in the complex plane). 
For $\mathbf{Re}\,z>0$,  
\be  
q_{a,b}(z)=-z^2\frac{b}{2}+\left(\delta+\frac{1}{2}\right)  
\ln(z^2+1)\pm \mathrm{i}{z}\mp\frac{1}{2}\ln\frac{1+\mathrm{i}z}  
{1-\mathrm{i}z}.  
\ee  
The saddle points are determined by the zeros of the derivative
$q_{a,b}^\prime(z)$,  
\be  
q_{a,b}^\prime(z)=-b z+\frac{(2\delta+1)z}{z^2+1}\pm \mathrm{i}\mp  
\mathrm{i}\frac{1}{z^2+1}{=}0.  
\ee  
The location of the solutions on the complex $z$--plane depends on the  
values of $\delta$ and $b$: there are actually three distinct regions in  
the ($b$\,--\,$\delta$) plane (see the left hand side of
figure \ref{f:cont}). We listed in 
table \ref{t:sol} the  solutions of the saddle equations which
correspond to the different values of  ($b$, $\delta$). In the table
$z_1$, $z_2$ denote  
\be  
 z_1=\frac{1}{2b}\left[\mathrm{i}+\sqrt{8b\delta-(2b-1)^2}   
\right],\quad  
z_2=\frac{\mathrm{i}}{2b}\left[1-\sqrt{(2b-1)^2-  
8b\delta} \right]. \label{e:z1z2}  
\ee    
\begin{table}[!t]  
\begin{center}  
\begin{tabular}{lccc}  
 & ($b,\,\delta$)\,$\in$\,I & ($b,\,\delta$)\,$\in$\,II &   
($b,\,\delta$)\,$\in$\,III\\  
\cline{2-4}  
$z_a:$ &$\left\{0,z_1\right\}$       &$\left\{0,z_2\right\}$,\,\,  
$(\left|z_2\right|<1)$& $\left\{0,z_2\right\}$,\,\,$(\left|z_2\right|>1)$ \\  
$z_b:$ & $\left\{0,z_1^\star\right\}$ &  
$\left\{0,z_2^\star\right\}$,\,\,  
$(\left|z_2\right|<1)$& $\left\{0,z_2^\star\right\}$,\,\,
$(\left|z_2\right|>1)$  
\end{tabular}  
\caption{The location of saddle points of $q_a$, $q_b$ are $z_a$,  
$z_b$ correspondingly. The $z_1$ and $z_2$ are given in (\ref{e:z1z2}).    
\label{t:sol}}  
\end{center}  
\end{table}
It is interesting that, the  ``trivial''$z=0$ saddle point is present in  
each region. However, when one continues analytically
 also the integrand of $I_2$ into the complex  
plane one finds that  its contribution exactly 
cancels the contribution of the $z=0$ saddle  
point of  $I_1$.   
Hence it is enough to handle $I_1$ with its non trivial saddle
points. The integration contours which are displayed on the right hand
side of figure \ref{f:cont}
correspond to region I.

For the actual evaluation with help of the saddle point technique
let us recall that its general strategy tells us that we have to deform  
the integration contour of $I_1$ in order the integration path  
passes over the saddle point (see the r.h.s. of figure \ref{f:cont}) because the  
dominant contribution of the integral comes from the neighbourhood of  
the saddle point. Near the saddle point $q_{a,b}(z)$ can be  
approximated by their second order Taylor expansions around   
the saddle point,  
\be  
q(z)=q(z_{s})+\frac{1}{2}\left.q^{\prime\prime}\right|_{z_s}   
\left(z-z_s\right)^2=q(z_{s})-D\frac{\tau^2}{2},  
\ee    
where $z_s$ denotes the corresponding saddle point and   
\be  
  D=\left|q^{\prime\prime}(z_s)\right|,\quad \,\, 
z=z_s+\tau\exp\left[\frac{-\mathrm{i}}{2}\left(\mathrm{arctan}  
\left(\frac{\mathbf{Im\,}  
        q^{\prime\prime}(z_s)}{\mathbf{Re\,}  
        q^{\prime\prime}(z_s)}\right)\mp{\pi}\right)\right],   
\quad \tau\in\mathbf{R}\,.  
\ee  
Practically speaking, the above form of $q_{a,b}$ provides a   
parameterization ($z(\tau)$) of the integration path near the saddle point in  
which the argument of the exponential function have constant imaginary  
part along the parameterized integration contour (stationary  
phase). Using the above parameterization for   
an integral of type $I_1$, one can write  
\bea  
 &&\int_0^\infty \mathrm{d} z \sqrt{z} \exp\left(\mu  
  q(z)\right)\approx \rme^{\mu q(z_{s}) } \int_{z_s} \mathrm{d} \tau  
\frac{\mathrm{d} z}{\mathrm{d} \tau}  \sqrt{z(\tau)} \exp\left(-\mu  
  D\frac{\tau^2}{2}\right) \approx  \nonumber  
\\  
 &&\approx \sqrt{z_s}\left.\frac{\mathrm{d}  
    z}{\mathrm{d} \tau}\right|_{z_s} \rme^{\mu q(z_{s})  
}\int_{-\infty}^\infty \mathrm{d} \tau  \exp\left(-\mu  
  D\frac{\tau^2}{2}\right) =\sqrt{z_s}   
\rme^{\mu q(z_{s}) }\sqrt{\frac{-2\pi}{\mu q^{\prime\prime}(z_s)}}, 
\label{e:steep}
\eea
where in the first line the notation  $\int_{z_s}$ means that the  
integration range is a small interval near $z_s,\, (\tau=0)$. In the  
second line this range is extended to $[-\infty,\,\infty]$ because the  
contribution to a Gaussian integral is small far away from the saddle  
point when $\mu$ is large. In addition we write $\sqrt{z_s}$ instead  
of $\sqrt{z}$ since it is a slowly varying  function near the saddle  
point. Higher order terms of the Taylor expansion   
of  $\sqrt{z}$ give $\mathcal{O}(1/\mu)$ corrections.  
    
It can be seen from (\ref{e:steep}) that the order of magnitude of  
the integral $I_1$ is determined by the real part of   
$q_{a,b}(z_s)$, which can be written for the regions $i=$\,(I, II,  
III) in the form $q^{i}_{a}(z^i_s)=R^i+\mathrm{i}\Gamma^i$,  
$q^{i}_{b}(z^i_s)=R^i-\mathrm{i}\Gamma^i$. After some algebra, where
one pays attention to the branch cuts of $\exp(\mu q(z_s))$ appearing in 
(\ref{e:steep}) one can obtain the following explicit expressions for the 
real and imaginary parts: 
{\allowdisplaybreaks   
\bea  
  R^{I}\!\!&=&\!-\frac{1}{4b}+\frac{b-1-2\delta}{2}+\frac{2\delta+1}{2}  
\ln\left(1+\delta\right)+\delta\ln\sqrt{\delta^2+\delta},\\  
  R^{II}\!\!&=&\!-\ln(2)-\frac{1}{4\beta}\left(2b-1-\sqrt{(2b-1)^2-8  
b\delta} \right)+\nonumber\\  
 &+&\!(\delta+1)\ln\left(2b+1-\sqrt{(2b-1)^2-8b\delta}\right)
-\delta\ln\left(2b-1-\sqrt{(2b-1)^2-8b\delta}\right)\!,\\  
  R^{III}\!\!&=&\!-\ln(2)-\frac{1}{4b}\left(2b-1+\sqrt{(2b-1)^2-  
8b\delta} \right)+\nonumber\\  
  &+&\!\!(\delta+1)\ln\left(2b+1+\sqrt{(2b-1)^2-8b\delta}\right)\!-\!  
\delta\ln\left(\left|2b-1+\sqrt{(2b-1)^2-8b\delta}\right|  
\right)\!,\\  
  {I^{I}}\!\!&=&\!\frac{\sqrt{8b\delta-(2b-1)^2}}{4b}+\frac{2\delta+1}  
{2}\mathrm{atan}\,\frac{\sqrt{8b\delta-(2b-1)^2}}{2b-1+  
4b\delta}\nonumber\\
 &-&\!\frac{1}{2}\mathrm{atan}\,\frac{\sqrt{8b\delta-(2b-1)^2}}  {2b-1-2b\delta}\,,\\  
 {I^{II}}\!\!& = &\!I^{III}=0.  
\eea  
}
Using these expressions together with the form of the normalization  
factor  $N$ from (\ref{e:N2}) and the exponential factor from  
 (\ref{e:i12}) one obtains that   
\be  
  0<R^{II,III}- b/2\approx \mathcal{O}(1),\quad\textrm{ therefore}
\quad N\exp\left(\mu(R^{II,III}- b/2 \right))\ll 1, 
\ee  
hence $R_n^\mu(\mu/2b)$ is exponentially small when $b,\,\delta$ are  
in regions II or III. Unlike in region I, where the exponential  
factors are   cancelled, and the result is  surprisingly simple:  
\be  
N^\prime \exp\left(R^{I}  
  -\mu\frac{b}{2}\right)\approx\left[\frac{\delta}  
{\delta+1}\right]^{1/4}. \label{e:ncanc} \\  
\ee    
Using (\ref{e:ncanc}) in (\ref{e:steep}) together with the explicit form of
$q^{\prime\prime}(z_s)$, one can obtain the expression (\ref{e:vegrn}). This
form of the asymptotic behaviour of the Laguerre polynomials is in
agreement with the results of \cite{Smith}, where asymptotic analysis of
the Laguerre polynomials was derived with help of the generating 
function of these functions.  \\

\end{document}